\documentclass[prc,aps,amsfonts,twocolumn,dvips,floatfix]{revtex4}
\usepackage{graphics,epsfig,amssymb}
\def\be{\begin{equation}}
\def\ee{\end{equation}}
\begin{document}

\title{ Accuracy of BCS-based approximations for pairing in small 
Fermi systems}

\author{N. Sandulescu$^{a,b}$ and G. F. Bertsch$^b$}
\affiliation{
$^a$ Institute of Physics and Nuclear Engineering, 76900 Bucharest, Romania \\
$^b$ DPTA/Service de Physique nucl\'eaire, F-91680 Bruy\`eres-le-Ch\^atel, France \\
$^c$ Institute of Nuclear Theory and Dept. of Physics, University of Washington, Seattle}

\begin{abstract}
We analyze the accuracy of BCS-based approximations for 
calculating correlation energies and odd-even energy differences
in 2-component fermionic systems with a 
small number of pairs.  The analysis is 
focused on comparing BCS and projected BCS treatments with the exact solution of
the pairing Hamiltonian, considering parameter ranges appropriate for 
nuclear pairing energies. We find that the projected BCS is quite
accurate over the entire range of coupling strengths in spaces of 
up to about $\sim 20$ doubly degenerate orbitals.   It is also quite 
accurate for two cases we considered with a more realistic Hamiltonian,  
representing the nuclei around $^{117}$Sn and  $^{207}$Pb. However, 
the projected BCS significantly  underestimates the energies for much larger 
spaces when the pairing is weak. 
\end{abstract}

\maketitle

\section{Introduction}

A theory of nuclear pairing based on the BCS approximation was considered  
for the first time 50 years ago \cite{bohr}.   
Since then, BCS-based approximations or more generally the Hartree-Fock-Bogoliubov
equation have become ubiquitous for calculating nuclear energies in the
framework of density functional theory or self-consistent mean-field theory.
It is thus important to understand the limitations of these approximations and
use more accurate theory when needed.  
In particular, the BCS Ansatz of a condensate with indefinite particle number becomes 
problematic in finite systems with weak pairing, as is the case for 
nuclei at shell closures.  Several ways have been proposed to improve
the theory\cite{rs}, beginning with number-projected BCS (PBCS) first proposed
by Bayman \cite{bayman} and Blatt \cite{blatt}.   It is our aim here
to evaluate the PBCS by testing it in situations for which an exact 
solution is available.  

To investigate the accuracy of PBCS approximation we shall consider a finite number 
of spin-$1/2$ fermions, e.g., neutrons or protons, distributed in a sequence of 
single-particle levels and interacting through a pairing force. We will mainly consider
the reduced BCS Hamiltonian given by 
\be
H= \sum_i^\Omega \varepsilon_i (a^\dagger_i a_i +a_{\bar i}^\dagger a_{\bar i})
  -  g \sum_{i, j}^\Omega a^\dagger_i a^\dagger_{\bar{i}} a_{\bar{j}}a_j.
\ee
Here $g$ is the strength of the pairing force acting in a space of $\Omega$ two-fold 
degenerate orbitals with the single-particle energies $\varepsilon_i$.

The Hamiltonian (1) is exactly solvable \cite{richardson1,richardson2} and it was used
in 60ies to make a critical analyses of the BCS approximation in finite Fermi 
systems. Thus in Ref.\cite{richardson2} Richardson studied the exact and the BCS
solutions of the Hamiltonian (1) with $\varepsilon_i=i$ and for systems with 
$\Omega = 8 - 32$ at half filling, i.e., with the number  of particles $N= \Omega$.  
Such systems plausibly model the pairing in deformed nuclei with  the active nucleons 
(for pairing calculations) filling the major shells 8-20, 50-82 and 82-126.  
The main conclusion of Ref. \cite{richardson2} was that BCS model strongly 
underestimate the pairing correlations even for relatively large values of the pairing strength. 
The question we address in this study is how much one could improve the BCS results 
relative to the exact model if we perform PBCS calculations.   Some of the issues
analyzed in this paper where also discussed recently in relation to metallic
grains studies \cite{duke,grains,delft}.  There has also been a recent
study in the nuclear physics context \cite{dussel}.  These authors found
a significant difference between the exact solution and the PBCS. In Sec. II 
we shall argue that the PBCS approximation is nevertheless quite accurate if it is 
applied in a limited window around the Fermi level of the order of one major shell 
in atomic nuclei.
  
Unfortunately, Richardson's model requires that the interaction matrix elements 
be equal in the pairing Hamiltonian \cite{footnote}.  It is essential to be able to treat 
the most general form of the Hamiltonian (1), with the  matrix elements computed as 
integrals over an effective two-particle potential, if the theory is to be a global 
one describing the entire nuclear mass table. In that case the Hamiltonian has a more
general form 
\be
\label{vij}
H= \sum_i^\Omega \varepsilon_i (a^\dagger_i a_i +a_{\bar i}^\dagger a_{\bar i})
  -  \sum_{i > j}^\Omega v_{ij} a^\dagger_i a^\dagger_{\bar{i}} a_{\bar{j}}a_j ,
\ee   
where $v_{ij}=v_{i\bar{i}j\bar{j}}$ are calculated with some two-body interaction
such as in Eq.(16) below. There is no algebraic solution for this more general case, 
but we can obtain useful results up to and beyond $\Omega=16$ using ordinary numerical 
matrix methods.  Some examples will be treated in Sec. \ref{CI}. The realistic calculations 
show that the PBCS approximation gives accurate results, confirming the conclusions based 
on Hamiltonian (1) and Richardson model. The fact that PBCS can provide a good description 
for the ground state of realistic Hamiltonians can be also seen from the large overlaps 
betweent the PBCS and the shell model wave functions \cite{ns97}.
 
 A good agreement between PBCS and the exact Richardson's solution we also find 
 for the occupation probabilities of single-particle levels, analysed in Sec. IV.
 It is worth to emphasize that this agreement is obtained with two 
 different wave functions for the ground state of the systems, i.e., a condensate
 formed by identical Cooper pairs in PBCS model, a non-condensate structure
 based on  distinct pairs in the case of exact Richardson's solution. These differences
 manifest clearly in the collectivity of the Cooper pairs, discussed in Sec. IV.
 The reason why such differences do not affect significantly the correlation energies
 and the occupation probabilities when the calculations are done in limited window
 around the Fermi level is not yet clear \cite{sambataro}.

Finally we would like to mention that another methodology  that is widely used to 
go beyond the BCS theory is the Lipkin-Nogami approximation.  We do not consider 
it here for following reasons.  First, it has been thoroughly studied in the past
and its strengths and deficiencies are well known.  It has a serious
shortcoming in that the approximation is not reliable near 
closed shell nuclei\cite{do93}. Since we seek approximation methods that 
cover all the extremes that arise in the nuclear mass table,  we
find this method unsuitable.

\section{Solutions of the BCS Hamiltonian}

The Hamiltonian (1) certainly describes the basic features of nuclear
 pairing correlations, and it is commonly solved using
the BCS or PBCS approximation.  Both BCS and PBCS methods are variational in that the approximation
is made on the wave function, and the energy is calculated as an 
expectation value.  The usual form of the BCS wave function is 
the well-known expression  $\Pi_i^\Omega (u_i + v_i a^\dagger_ia^\dagger_{\bar
i})|\rangle$, but for putting it in the context of the other treatments one
can write it as an exponentiated product of a pair operator,
\begin{equation}
\Gamma^\dagger = \sum_i x_i a^\dagger_i a^\dagger_{\bar{i}} .
\end{equation}
The BCS ground state can then be expressed as 
a coherent superposition of pairs, i.e.,
\begin{equation}
| BCS \rangle \propto  e^{\Gamma^\dagger} | 0 \rangle \equiv  \sum_n
\frac{(\Gamma^\dagger)^n}{n!} | 0 \rangle . 
\end{equation}
The mixing amplitudes of the pair operator, written usually as $x_i=v_i/u_i$, are given by the
well-known BCS equations. 
The PBCS approximation is obtained by restricting the expansion in Eq.(3) to the term having the 
required number of particles. Thus, in PBCS  the ground state wave function 
can be expressed
\begin{equation}
\label{pbcs.wf}
| PBCS \rangle \propto (\Gamma^\dagger)^{N_{pair}} | 0 \rangle,
\end{equation}
where $N_{pair}$ is the number of pairs. The PBCS equations, which determine the mixing amplitudes $x_i$ of
the pair operator (2), are derived by minimizing the average of the Hamiltonian in the state (5).
They can be solved by using the residual integrals technique \cite{pbcs}. Alternatively,
if the number of pairs is not too large, the amplitudes $x_i$ can be found by using the
technique of recurrence relations.  

As shown in Ref. \cite{richardson1}, the pairing Hamiltonian (1) can be solved 
exactly.  The solution resembles eq. (\ref{pbcs.wf}) except that the 
operator $\Gamma^\dagger$ is replaced by $N_{pair}$ different
pair operators $B_\nu^\dagger$,
\begin{equation}
| \Psi \rangle = \prod_\nu^N B^\dagger_\nu | 0 \rangle.
\end{equation}
The pair operators have the form
\begin{equation}
B^\dagger_{\nu} = \sum_i \frac{1}{ 2\varepsilon_i-E_{\nu} } a^\dagger_i
a^\dagger_{\bar{i}} . 
\end{equation}
They depend on energy parameters $E_\nu$ obtained by
solving the set of nonlinear equations
\begin{equation}
\frac{1}{g} - \sum_j \frac{1}{2\varepsilon_j-E_\nu}
+\sum_{\mu \not= \nu} \frac{2}{E_\mu-E_\nu}
 = 0 .
\end{equation}
The sum of pair parameters $E_\nu$ gives the total energy of the system, i.e.,
\begin{equation}
E = \sum_\nu E_\nu .
\end{equation}

In the limit $g=0$ the pair energies $E_\nu$ of the ground state
solution coincide with the lowest single-particle energies, i.e.,
$E_\nu=2\varepsilon_\nu, (\nu=1,2, ...N_{pair})$.
When the interaction is turned on, the pair energies evolve toward lower
values and could become complex two at a time. This fact was used by
Richardson to obtain a set of equations in which the singularities are
removed \cite{richardson2}.

For small values of $g$ in finite systems, the BCS condensate collapses and
the BCS approximation gives zero correlation energy.  On the other hand, 
due to the finite distance between the levels, for small values of the interaction 
strength the pairing interaction energy can be calculated perturbatively. 
This is opposite to what happens in finite systems where the pairing correlations 
depends exponentially on pairing strength in the weak coupling limit. The perturbative 
solution can be easily derived by enumerating the two-particle two-hole configurations 
starting from the lowest energy configuration, or by taking the small-$g$ limit of the
Richardson equations \cite{grains}. 

The second-order perturbation result for the interaction energy, i.e., the energy gained by the 
system when the interaction is turned on, is given by
\be
\label{pert}
E_{corr}^P= 
{g^2\over 2} \sum_{i=1}^{N_{pair}} 
\sum_{j=N_{pair}+1}^{\Omega} 
\frac{1}{\varepsilon_j - \varepsilon_i} 
\ee
This expression is valid for even $N$; for odd $N=N_{pair}+1$ the $j$ sum
begins at $j=N_{pair}+2$.
It can be shown \cite{grains} that the approximation (\ref{pert}) for the interaction energy 
is valid if $g<g_P \equiv g^*(1-g^*)$, 
where $g^*$ is the convergence radius of the perturbative expansion given by 
\begin{equation}
\frac{1}{g^*} = \sum_{j=1}^{N_{pair}} \frac{1}{\varepsilon_j}.
\end{equation}
 
\subsection {Correlation Energies and Pairing Gaps} 

\begin{figure}
\begin{center}
\includegraphics*[scale=0.30,angle=-90]{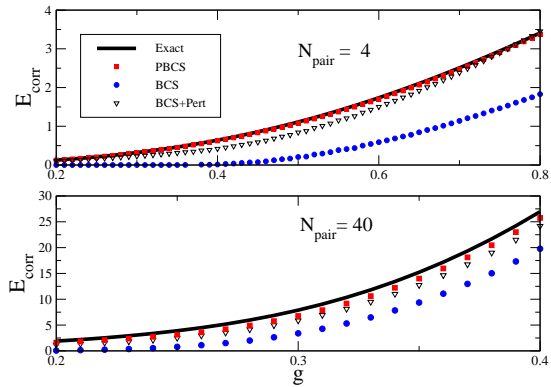}
\caption{ Correlation energies for the Hamiltonian (1), calculated
in various approximations.  Upper panel: $\Omega=8, N_{pair}= 4$; lower
panel: $\Omega=80, N_{pair}= 40$.} 
\end{center}
\end{figure}

\begin{figure}
\begin{center}
\includegraphics*[scale=0.30,angle=-90]{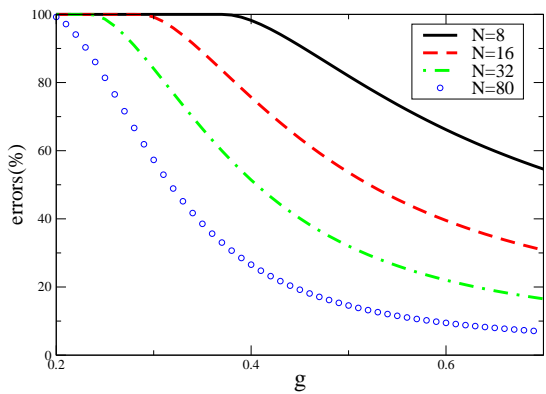}
\caption{ Errors of the BCS approximation for correlation energies.}
\end{center}
\end{figure}

\begin{figure}
\begin{center}
\includegraphics*[scale=0.30,angle=-90]{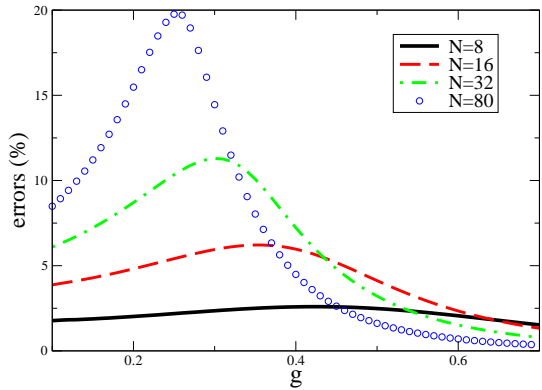}
\caption{ Errors for the correlation energies calculated in the PBCS
approximation.}
\end{center}
\end{figure}

\begin{figure}
\begin{center}
\includegraphics*[scale=0.30,angle=-90]{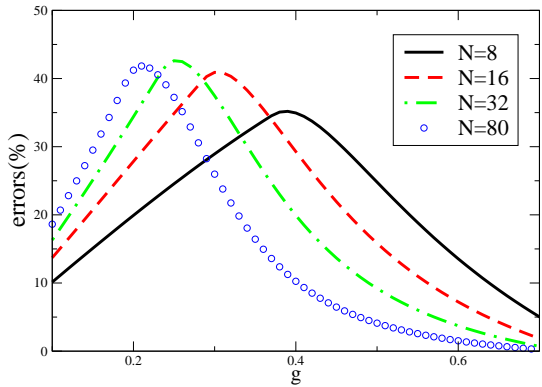}
\caption{ Errors for the correlation energies approximated as the 
sum of the BCS energy and the second-order perturbative energy, 
Eq. (\ref{sum}). }
\end{center}
\end{figure}

In this section we shall discuss the correlations energies and the pairing gaps
provided by the Hamiltonian (1) for a single-particle spectrum formed by 
$\Omega$ two-fold degenerate orbits  with uniformly spaced energies, i.e., 
$\varepsilon_i=i$. The analyses is done for systems formed by N particles 
distributed in $\Omega=N$ levels with  $\Omega = 8 -80 $. For the strength
of the interaction we use the range $g = 0.1 -0.8$, which covers all the
interesting coupling regimes met in nonspherical atomic nuclei. In all figures discussed
in this section the energies, the pairing gaps and the interaction strength
are given in units of single-particle levels spacing. 

We first discuss the correlation energies calculated exactly and in various 
approximations. The correlation energies are defined by the difference in energies between 
a single Slater determinant and the pair-correlated state, i.e.,
\be
E_{corr}(g) = E_{HF}  - E(g).
\ee
The results are shown in Fig.1. From this figure is obvious that the 
BCS seriously underpredicts the 
correlation energy while the PBCS does much better.  To see the
differences more quantitatively, in the next figures we show for
each approximation how the error depends on $g$ and $\Omega$.  The BCS
errors shown in Fig. 2 are large, making this approximation completely
unreliable. The results for the PBCS approximation are shown in 
Fig. 3.  The errors are much smaller, but become unacceptably large
for the biggest space, $\Omega=80$.  In spacing corresponding to 
a single major shell the error is well under 10\%.  
It is important to emphasize that the example discussed above does not
contradict the fact that pairing models become more accurate if one moves
toward the thermodynamical limit \cite{richardson3}. To reach this limit
the particle number is increased but, in the same time, the calculations are
done keeping a fixed energy window around the Fermi energy. In this way the
increase of particle number has as effect an increase of level density
around the Fermi level which, in turn, is increasing the pairing
correlations and by that the accuracy of pairing models. This is different
from what happens in the system with 40 pairs discussed above and in many
BCS and HFB calculations performed for atomic nuclei, where the energy window around
the Fermi level is increased such that to include the deep bound nucleons.
As discussed above, by this procedure the accuracy of pairing calculations
becomes worse not better.

 
The question which arises is why BCS approximation strongly underestimates the
pairing correlations in finite systems. This question was recently addressed in 
relation to metallic grains calculations. Thus, in Ref.\cite{grains} it is
argued that BCS works properly only for so-called "condensed" levels, i.e., the 
levels with the energies located in the interval $I=|\Delta-\mu |$, where 
$\Delta$ is the
gap parameter in the BCS equations and $\mu$ is the Fermi energy. This conclusion is supported by the 
observation that the correlation energy calculation in BCS is close to the exact 
result obtained if from the exact solution is kept only the contribution of 
condensed levels (in the exact solution the condensed levels have usually complex 
pair energies). On the other hand, it was found that the contribution 
of levels located outside the interval $I$ are underestimated in BCS. Since 
in the exact solution the Cooper pairs corresponding to these levels have the 
pair energies close to the single-particle levels, one expects that the 
contribution of these levels to pairing correlations could be treated 
perturbatively. Based on these observations it was found \cite{duke,grains,delft} 
that in metallic grains the correlation  energy could  be approximated by a formula 
combining the BCS and the perturbative expressions in the sum
$E_{corr}^{BCS}(g)+\Delta+a(g) E_{corr}^P$,
where the last term is the perturbative result and $a(g)$ is a function of
the order of unity determined by the fitting protocol. The BCS contribution
is approximated by the first two terms representing the condensation energy
and the pairing gap for infinite system. The later accounts for the size
correction to the bulk result (first term). In order to see if such an
approximation could work for the small systems analyzed here, 
we have simply replaced the first two terms in the above expression by the BCS result (for
finite systems) and we take $a(g)=1$, 
\be
\label{sum}
E_{corr} = E_{corr}^{BCS} + E_{corr}^P.
\ee
The corresponding results are shown in Fig. 1 by inverted triangles and the corresponding
errors are given in Fig. 4. It can be seen that this simple approximation 
works surprisingly well for a wide range of the pairing strengths,
including the physical region up to $g \sim 0.8$.  Of course, it is not
applicable to situations where there is a degeneracy in the single
energies at the Fermi level, since perturbation theory diverges in that
situation.

\begin{figure}
\begin{center}
\includegraphics*[scale=0.30,angle=-90]{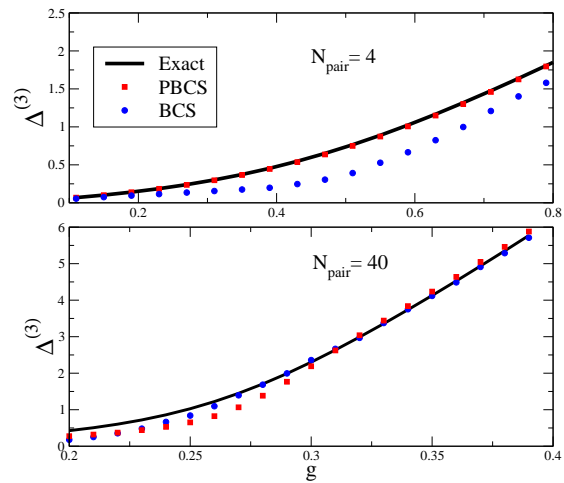}
\caption{ Pairing gaps (Eq.~(\ref{gap})) calculated in various approximations.
Upper panel:   $\Omega=8, N_{pair}= 4$; lower
panel: $\Omega=80, N_{pair}= 40$.}
\label{gaps}
\end{center}
\end{figure}
\begin{figure}
\begin{center}
\includegraphics*[scale=0.30,angle=-90]{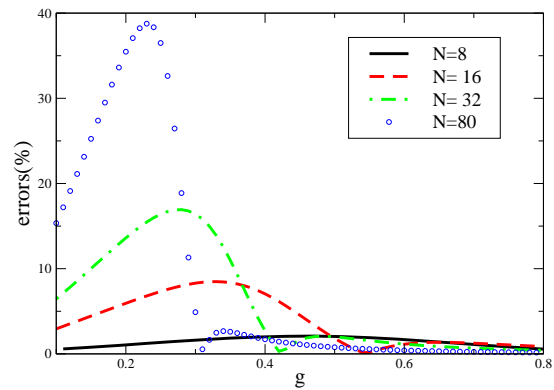}
\caption{ Error in pairing gaps calculated with the PBCS approximation.
}
\label{ergap_pbcs}
\end{center}
\end{figure}

We next compare the pairing gaps calculated in the BCS and PBCS approximations with
the exact values. The gap is defined as the second difference of energies between 
an odd-$N$ system and its 
neighbors with even-$N$,
\be
\label{gap}
\Delta^{(3)}(N) = {1\over 2} \left(2  E(N) - E(N-1) - E(N+1)\right).
\ee
We shall call $\Delta^{(3)}(N)$ the gap at number $N$. 
The results at the smallest and largest spaces are 
shown in Fig. \ref{gaps}.
One can see that 
PBCS gives accurate gaps in all coupling regimes, contrary to the 
larger systems analyzed in metallic grain studies \cite{duke,grains,delft}.
The errors associated with the PBCS approximation are shown in more detail
in Fig. \ref{ergap_pbcs}.  It is probably acceptable to tolerate an error
up to 10\% for the gap, but not higher.  This confirms for the gaps that
the PBCS is only reliable up to moderate size spaces.

It is interesting to note that in spite of the large errors in the
correlation energies, in the physical region of the strength parameter the
BCS pairing gaps come much closer to the exact results. However,
contrary to the PBCS results, the good agreement of the BCS gaps to the exact
values in the region of well-developed pairing correlations is just a
manifestation of the errors cancellation when the subtraction is performed
in equation (14). Consequently, fixing the pairing force by the odd-even
mass difference, as usually done in nuclear structure calculations, does not
guarantee a good description of correlation energies within the BCS
approximation.

\section {PBCS for realistic pairing interactions}
\label{CI}

Ultimately, the theory of nuclear binding energies should be based on 
realistic interactions dropping the constant-$g$ approximation. The 
Richardson solution 
has been somewhat generalized to encompass separable pairing
interactions\cite{ba07},
but to be truly realistic the Hamiltonian must allow completely general
interactions in the many-body space of pairwise occupated 
orbitals.  These Hamiltonians can be solved by straightforward
configuration interaction (CI) methodology, in which a Hamiltonian matrix
is constructed in the Fock space of the orbitals and diagonalized by
standard linear algebra operations.  The size of the space $D$ needed to
represent the most general paired wave function is given by the number
of combinations of $N_{pair}$ orbitals out of total of $\Omega$,
\be
D= \left(\begin{array}{l} \Omega \\ N_{pair} \end{array}\right).
\ee
For example, for $\Omega=16$ orbitals and $N=16$ particles the 
dimension of the space is 12,870.  The lowest eigenvector for a space
of this size is easily calculated on serial computers using the 
Lanczos algorithm.

We have carried this out for two examples in which the pairing is 
very different.  The Hamiltonian makes use of the 
orbital energies and wave functions from the global mean-field
calculations
of Ref. \cite{be06}, which are based on the Skyrme SLy4 energy functional.
Pairing is strong in the Sn isotope chain, and we will take
$^{117}_{67}$Sn$_{50}$ and its neighbors as an example of where the
pairing is well developed.  The second example is $^{207}$Pb.  In the global systematics
of neutron pairing gaps, the one at $^{207}$Pb is the smallest
($\Delta^{(3)} = 0.32$ MeV), so this provides a good test of the 
approximation methods in the weak pairing limit.

We use the published code  ev8 \cite{cpc} to recalculate the needed orbital
properties, starting from the wave functions provided in the original global
survey \cite{be06}.  The orbitals are represented internally
in the code with a
3-dimensional mesh, so they need not have good angular momentum
quantum numbers.  However, for the even-$N$ Sn and Pb isotopes, the mean
field solution is spherically symmetric and the orbitals can be given
angular momentum assignments.  In Ref. \cite{be06} the pairing interaction
was taken as a density-dependent contact interaction in a space truncated to
a band of width 10 MeV about the Fermi energy.  Here we use an ordinary
delta function to generate the pairing matrix elements,

\be
\label{delta}
v_{ij} = v_0\int d^3 r |\phi_i(r)|^2 |\phi_j(r)|^2,
\ee
where the $\phi(r)$ are orbital wave functions. 
The strength $v_0$ has been fitted to the global systematics of 
pairing gaps \cite{be08} and the self-consistent orbitals were generated
with that value.  The single-particle
energies and the matrix of $v_{ij}$ values were then used as input data
for separate codes to solve the Hamiltonian Eq.~(\ref{vij}).  We
calculate the correlation energies and the pairing gaps using 
a range of values for $v_0$ obtained by scaling the matrix elements
obtained from the ev8 code.

\begin{figure}
\begin{center}
\includegraphics*[scale=0.3,angle=-90]{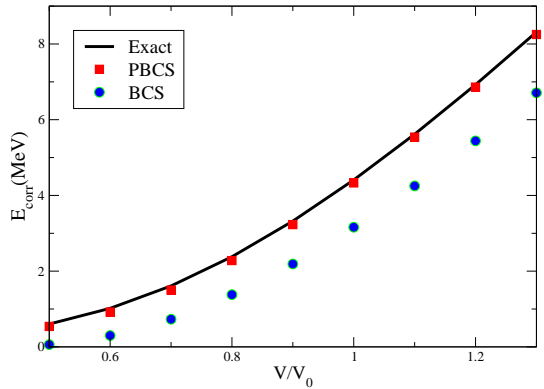}

\caption{\label{sn116}
Pairing correlation energy in $^{116}$Sn. The pairing
interaction is a delta function with a strength scaled from the
nominal value $v_0=465$ MeV-fm$^3$ by a factor given on the
abscissa.  The orbital space is the 16 orbitals around the Fermi energy
as described in the text.  Solid line shows the result of exact diagonalization.}
\end{center}
\end{figure}

For the Sn isotopes with neutron number $N$ around
68, there are $\Omega=16$ orbitals in the 10 MeV window, originating from the
$d_{5/2},g_{7/2},s_{1/2},d_{3/2}$ and $h_{11/2}$ shells of the spherical
shell model.  This space is small enough to permit 
the exact calculations to be performed without
special numerical difficulties.
Fig.~\ref{sn116} shows the neutron correlation energy in $^{116}$Sn as a function 
of interaction strength $v_0$.  The range for $v_0$ includes the value 
$v_0 \sim 450 $ MeV fm$^3$ that has been fitted to the global gap systematics
using the BCS approximation\cite{be08}.  The comparison in Fig.~\ref{sn116}
confirms the results obtained with the Hamiltonian with the constant-$g$ pairing.  
Namely, the BCS systematically underpredicts the correlation energy while the
PBCS tracks the exact energy very well.  PBCS describes also very well
the pairing gaps, as  shown in Fig. 8. There is also a good agreement
for the BCS gaps but, as already noticed in the previous section, this
agreement is in fact a manifestation of errors cancelation.

\begin{figure}
\begin{center}
\includegraphics*[scale=0.3,angle=-90]{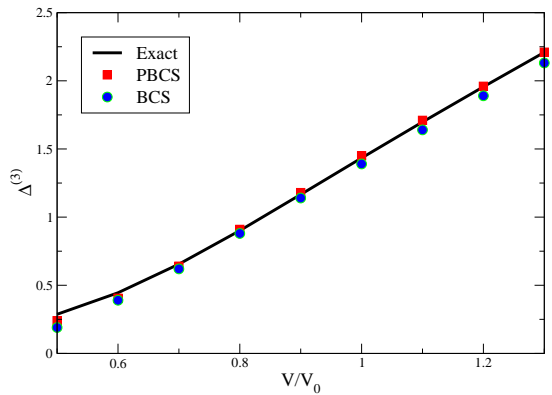}
\label{gap117}
\caption{Neutron pairing gap at $^{117}$Sn.  The correlations energies
for the three nuclei needed for Eq.~(\ref{gap}) were calculated with the
same functional and in the same space as in Fig. \ref{sn116}.
}
\end{center}
\end{figure}

\begin{figure}
\begin{center}
\includegraphics*[scale=0.30,angle=-90]{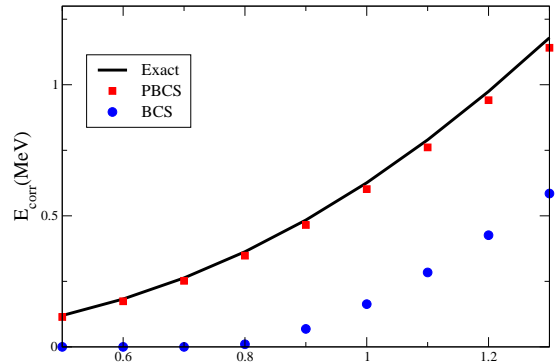}
\caption{\label{fig:pb206}
Pairing correlation energy in  $^{206}$Pb.  For these calculations, 
the space was truncated to $\Omega=16$ by including the 11 highest orbitals
below the $N=126$ magic number and the 5 $g_{9/2}$ orbitals above $N=126$.
This corresponds to an energy window of about 8 MeV .
}
\end{center}
\end{figure}

\begin{figure}
\begin{center}
\includegraphics*[scale=0.30,angle=-90]{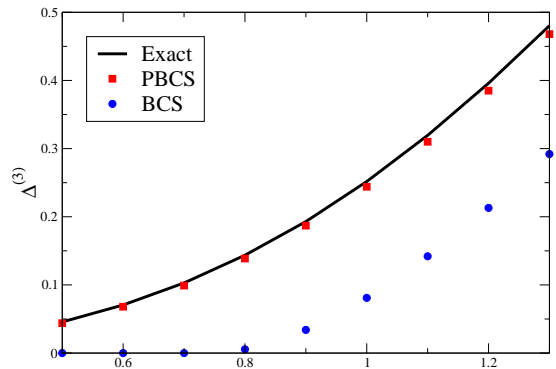}
\caption{\label{fig:pb207}
Neutron pairing gap at $^{207}$Pb.  For these calculations we used
the same space as in Fig. 9.}
\end{center}
\end{figure}

Next we show the results for Pb.  Here the energy truncation to a 10 MeV
window gives a space that is still too large to easily 
perform the exact calculation, so we truncated it further ($\sim 8$ MeV
window) to 
leave $\Omega=16$ orbitals.
Fig. \ref{fig:pb206} shows 
the correlation energy in this space and the BCS and PBCS
approximations.  One sees that the PBCS keeps an accuracy much better than
100 keV, while the BCS is off by more than a half of an MeV. 
The correlation energies in the other
nuclei needed for the $^{207}$Pb gap are very small and in BCS there is
no condensate in $^{207}$Pb and $^{208}$Pb.  As a result, the BCS error 
for the $^{206}$Pb
correlation energy is not well canceled in formula for the gap energy.  
Fig. \ref{fig:pb207} shows
the calculated gap in the three treatments.  One sees again that the PBCS
is remarkably accurate.  The BCS error of 100-200 keV is quite significant
on the scale of the empirical gap energies, which fluctuate around an
average of 1 MeV with an r.m.s. deviation of 300 keV.


\section {Occupation probabilities and two-body correlations}

To probe the accuracy of BCS-based models relative to the exact solution
we have also analyzed the occupation probabilities and the two-body 
correlations induced by the pairing force. The results discussed in this
section correspond to the Hamiltonian (1) with $\varepsilon_i=i$.  

\subsection{Occupation probabilities}

 The BCS occupation probabilities are calculated by solving the standard BCS equations
 (here we take into account the renormalisation of the single-particle energies by the
 diagonal term of the interaction) while the PBCS values are obtained by using the residual 
 integral method described in Ref. \cite{pbcs}.

In the exact solution the occupation probabilities $v_i^2$ are given by \cite{richardson2}
\begin{equation}
v_i^2= \sum_{\nu=1}^N \frac{a_\nu}{(2i-E_\nu)^2} ,
\end{equation}
where $a_\nu$ are obtained by solving the set of equations
\begin{equation}
[C_\nu -2 \sum_\mu \frac{1}{(E_\mu-E_\nu)^2} ] a_\nu
+2\sum_\mu \frac{a_\mu}{(E_\mu-E_\nu)^2} = 1 
\end{equation}
and $C_\nu$ are given by
\begin{equation}
C_\nu = \sum_{i=1}^{2N} \frac{1}{(2i-E_\nu)^2} .
\end{equation}
To solve the equations above is convenient to rewrite them in terms of the real and the imaginary parts 
of pair energies $E_\nu$. The corresponding expressions can be found in Ref. \cite{richardson2}.
 
For the discussions below we shall use the product between the occupation and non-occupation 
probabilities, i.e., $\kappa^2_i \equiv v_i^2(1-v^2_i)$, which provides relevant informations 
on the diffusivity of the Fermi  sea and the entanglement properties of pairing tensor (see
Eq.(21) below). 
\begin{figure}
\begin{center}
\includegraphics*[scale=0.30,angle=-90]{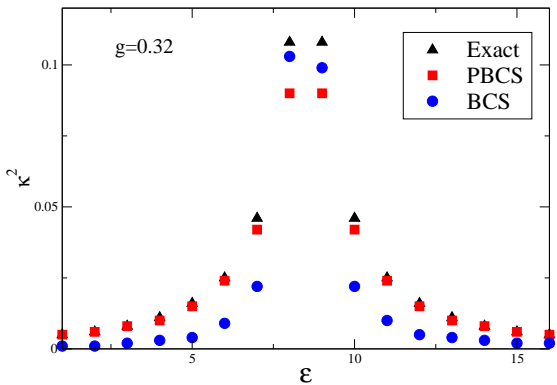}
\caption{Occupation factors $\kappa^2_i=v_i^2(1-v^2_i)$ for $N_{pair}=8$,
$\Omega=16$ and $g=0.32$. The results correspond to the Hamiltonian (1)
with $\varepsilon_i=i$.}
\end{center}
\end{figure}
\begin{figure}
\begin{center}
\includegraphics*[scale=0.30,angle=-90]{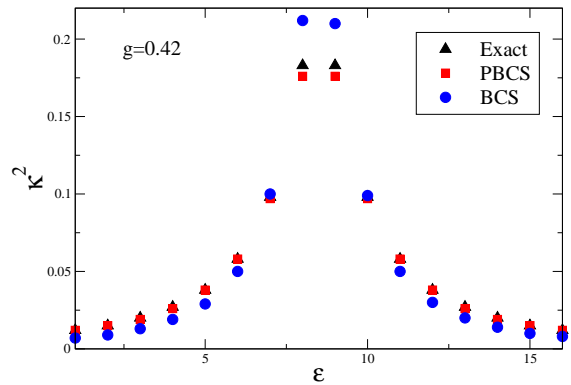}
\caption{The same as in Fig. 11 but for $g=0.42$}
\end{center}
\end{figure}
\begin{figure}
\begin{center}
\includegraphics*[scale=0.30,angle=-90]{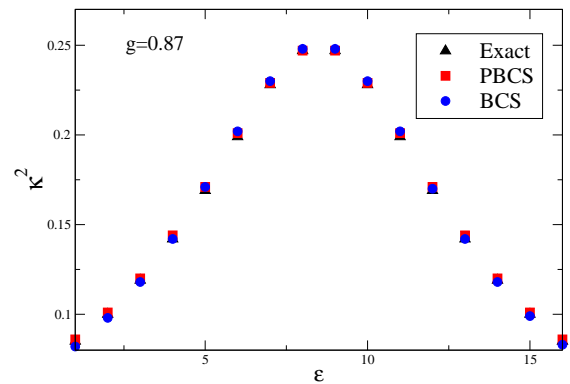}
\caption{The same as in Fig. 11 but for $g=0.87$}
\end{center}
\end{figure}

In Figs. 11-13 are shown the values of $\kappa^2_i$ for a system with N=8 pairs and for three values of 
the strength parameter corresponding to the weak  ($g=0.32$), intermediate ($g=0.42$) 
and strong coupling ($g=0.87$) regimes. For these strength values the gaps (exact results)
are approximatively equal to 0.6, 1.0, 5.0, respectively. As seen from Figs. 11-13, 
the values of $\kappa^2_i$ for the states outside the interval $I=|\Delta-\mu |$
are rather well described by  PBCS and underestimated by BCS.  
On the other hand, one can notice that for the states which are the closest to the 
chemical potential BCS gives in the weak and intermediate
coupling regimes larger values than PBCS and the exact solution. In the strong coupling
regime shown in Fig. 13, when all levels are inside the interval $I$, the BCS and 
PBCS results become close to the exact values for all levels included in the 
calculations. Hence, when the pairing correlations are well-developed, 
BCS describes rather well the occupation probabilities even though the errors 
for the condensation energies are large.

\subsection{Two-body correlations}

The distribution  of occupation probabilities around the Fermi level determines the intensity of
two-body correlations, which could be eventually probed in pair transfer processes. 
Two-body correlations are commonly introduced  by the two-body density defined by 
\begin{eqnarray*}
\rho_2(x_1,x_2)&=&
\sum_{\sigma_3 ..\sigma_N} \int
| \Psi(x_1,x_2,...,x_N) |^2 d\vec{r}_3...d\vec{r}_N\\ 
&=&
\langle0|a^+(x_1)a^+(x_2)a(x_2)a(x_1)|0\rangle ,
\end{eqnarray*}
where $\Psi$ is the many-body wave function, $x$ denotes the radial and spin coordinate,
i.e., $x \equiv \vec{r} \sigma$, and $a^\dagger(x)$ is the particle creation operator. 
To separate the genuine two-body correlations induced by 
pairing correlations from the correlations of Hartree-Fock type, the two-body density 
is usually written in the following form \cite{yang}
\begin{equation}
\rho_2(x_1,x_2) =
\rho(x_1)\rho(x_2)-|\rho(x_1,x_2)|^2 + |\kappa(x_1,x_2)|^2 ,                   
\end{equation}
where $\rho(x)$ is the (local) particle density while $\rho(x_1,x_2)$ 
is the non-local (exchange) part of particle density. The last term defines the genuine 
two-body correlations, i.e., the correlations not included in the independent mean-field
picture of fermion motion. In the BCS approximation the last term corresponds to the 
pairing tensor in the coordinate representation, i.e.,
\begin{equation}
\kappa(x_1,x_2) =
\langle0|a(x_2)a(x_1)|0\rangle = 
\sum_i \kappa_i \varphi_i(x_1) \varphi_{\bar{i}}(x_2) ,
\end{equation}
where $\kappa_i \equiv \langle 0|a_ia_{\bar{i}}|0\rangle=u_iv_i$ is the pairing tensor in a 
single-particle basis defined by the operators $a^\dagger_i$ and $\varphi_i$ are
the associated wave functions.

According to its definition, the pairing tensor in the coordinate representation provides 
information about the spatial correlations between two nucleons {\it irrespective} 
if they belong or not to the same Cooper pair. The spatial structure of these correlations
in atomic nuclei have been recently discussed in Refs. \cite{ns2005,np}. If we need to investigate
only the spatial correlations between  the nucleons belonging to the same pair, instead of pairing 
tensor one should analyse the pair wave function. The latter has a different structure in BCS-based models 
compared to the exact solution. Thus, in BCS and PBCS models all pairs are described by the same 
wave function   
\begin{equation}
\phi(x_1,x_2) 
= C \sum_i x_i \varphi_i(x_1) \varphi_{\bar{i}}(x_2) ,
\end{equation}
where the mixing amplitudes $x_i$ and the normalization factor $C$
depend on the approximation used to describe the condensate. 
Thus, in BCS $x_i=v_i/u_i$ and $C= \sum_i v^2_i/u^2$, where ${v_i,u_i}$ are the 
occupation amplitudes provided by the BCS equations. Formally, the same expressions 
can be used for the PBCS model but in this case the amplitudes ${v_i,u_i}$ 
are just variational parameters, not occupation amplitudes. 

For the exact solution each pair is described by a specific wave function, i.e.,
\begin{equation}
\phi_{\nu}(x_1,x_2) =
\sum_i \frac{1}{2\varepsilon_i-E_{\nu}}
 \varphi_i(x_1) \varphi_{\bar{i}}(x_2) .
\end{equation}
Therefore the correlations among the nucleons belonging to the same pair can be 
rather different according to the pair function they belong to.
\begin{figure}
\begin{center}
\includegraphics*[scale=0.30,angle=-90]{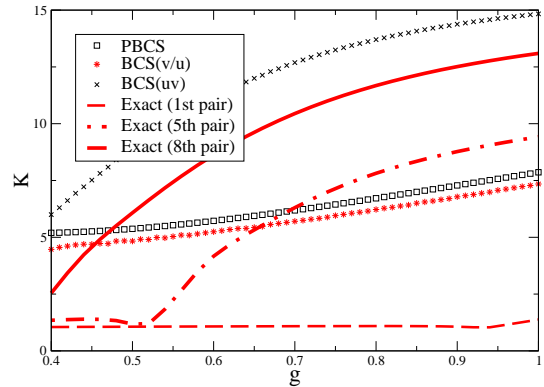}
\caption{The Schmidt numbers (Eq. 24) for a system with $N_{pair}=8$
and $\Omega=16$.}
\end{center}
\end{figure}
To characterize the amount of correlations in the pairing tensor (21) and in the pair wave 
functions (22-23) we employ the quantity \cite{grobe}
\begin{equation}
K=1/\sum_i w^4_i,
\end{equation}
where $w_i$ are the mixing amplitudes of the normalized two-body wave functions (21-23).
The quantity K is sometimes called the Schmidt number \cite{law} and it gives
a global indication of the degree of entanglement in two-body systems. Thus, the minimum 
value of K is 1 and is obtained when in the expansion (24) only one term is non-zero; 
this case corresponds to no entanglement since the two-body wave function is
split in the individual wave functions of the two- particle system. 
The maximum possible value of K is obtained when the weights $w_i$ have the same value 
for all terms in the expansion.

In Fig 14 is shown how K number evolves as a function of the strength parameter.
One can see that the entanglement properties of Cooper wave functions 
are rather similar in BCS and PBCS approximations and very different
from the entanglement of the correlation function (21). As already
stressed above, the latter describes the correlations between two 
generic fermions, which do not necessarily belong to the same 
Cooper pair.

 A particular behavior have the Cooper pairs (23) described by the 
exact solution. Their entanglement properties depends on how far are their pair 
energies $E_\nu$ (more precisely, their real part) from  Fermi 
level. Thus, as seen in Fig 14, the $8th$ pair (i.e., the one corresponding 
to $E_8$), which is the closest to Fermi level, is the most entangled. 
At the other extreme is the $1st$ pair, corresponding to $E_1$, which remains 
almost uncorrelated in all coupling regimes. An intermediate behavior has 
the $5th$ pair which remains uncorrelated  up to $g \approx 0.53$ and then 
its entanglement is increasing rather fast, in a similar fashion as for the 
$8th$ pair.  The strength value for which K number of the $5th$ pair starts 
to increase corresponds to the value at which the pair energy $E_5$ becomes complex. 
As mentioned previously, the pair energies become complex when the corresponding level 
(in the limit $g=0$) becomes condensed, i.e., enters in the interval $I=|\Delta -\mu|$. 
Hence, in the exact solution at a given value of the strength only some pairs become 
entangled, namely the ones corresponding to the condensed levels. This is very different from 
what happens in BCS and PBCS models in which all Cooper pairs are identical and therefore all of 
them have the same entanglement properties.

\section{Summary and Conclusions}

In  this paper we have studied how reliable are the approximations used to solve
the pairing Hamiltonian when the parameters are chosen in a range appropriate 
to calculate nuclear ground states.  As is well known, the BCS approximation does not
do well for correlation energies.  On the hand, we find that PBCS,  with
its variation after number projection, is highly accurate over most the 
interesting parameter range for moderate size orbital spaces.  This
applies to orbital spaces such as a single major shell or energies
truncations of the order of 5 MeV around the Fermi level.  Much larger
spaces, for example including all the occupied orbitals, give a degradation in
the performance of the PBCS that we do not yet understand. However, by renormalizing
the pairing interaction, the calculations in large spaces could be reduced
to smaller ones in which the PBCS gives better results.  One way the renormalization
could be done is to demand the same computed gaps when the number of states available 
for the active pairs is changed. For example,  as seen in Fig. 3, if in the PBCS calculations 
we decrease the number of active pairs from 40 to 8 and, in order to keep the same gap, we increase 
the strength from 0.27 to 0.5, the error in the correlation energy drops 
from $20\%$ to below $5\%$. Hence, to reduce the errors of PBCS calculations 
is preferable to restrict the calculations to a small number of active particles 
with energies located close to the Fermi level.  In any case,
single-major shell truncations are widely used and there is no reason
to not use the PBCS with those conditions.

Both the BCS and PBCS seem to do well on calculating pairing gaps with
the the reduced BCS Hamiltonian Eq. (1).  However, in the case of BCS the
improvement arises from a cancellation of errors.  The BCS underpredicts
the correlation energies, but more seriously is missing the true values
for odd systems.  In the realistic case of $^{207}$Pb the errors did
not cancel well, and so we regard the BCS as unreliable at the level
of 100 keV or so.  The PBCS maintained its accuracy for the two
realistic cases we considered, confirming our overall assessment of
its reliability.

 We have also found that PBCS describes accurately the occupation
 probabilities of the single-particle levels. However, contrary to 
 the agreement found for the correlations energies and the occupation
 probabilities, the entanglement properties of pair wave functions are 
 very different in PBCS and in the exact solution.  Thus, while in PBCS the
 entanglement of Cooper pairs depends smoothly on the interaction strength, 
 in the exact wave function, formed by non-identical pairs, the entanglement
 properties of Cooper pairs depend strongly on the position of their pair 
 energies relative to chemical potential.

Finally, we would like to mention that a ground state based on non-identical
pairs, specific to the exact solution, could be more appropriate for the
description of loosely bound systems such as nuclei close to the drip lines.
For such nuclei one expects that the properties of Cooper pairs formed by
the valence nucleons moving in loosely bound and continuum single-particle
state to be rather different from the pairs formed by the deeper bound
nucleons.

\vskip 0.5cm
\noindent
{\bf Acknowledgments}
\vskip 0.2cm
\noindent
We thank Jorge Dukelsky and Stuart Pittel for valuable discussions and for
their assistance in solving the Richardson and PBCS equations.  
This work was supported by the UNEDF SciDAC Collaboration under DOE
grant DE-FC02-07ER41457, by DE-FG02-00ER41132 and by the grant IDEI
nr. 772.


\begin{thebibliography}{99}
\bibitem{footnote} This requirement can be relaxed under certain 
conditions. See refs. \cite{gaudin}, \cite{ba07} and \cite{pa98}.
\end{thebibliography}
\end{document}